\documentclass[aps,prb,twocolumn,superscriptaddress,showpacs,floatfix]{revtex4}
\usepackage{graphicx}
\begin{document}
\title{Raman and Infra-red properties and layer dependence of the phonon
  dispersions in multi-layered graphene}

\author{Jin-Wu Jiang} \email[Electronic address:]{jwjiang@itp.ac.cn}
\affiliation{Institute of Theoretical Physics, Chinese Academy of
  Sciences, Beijing 100080, China }
\author{Hui Tang} \affiliation{Institute of Physics, Chinese Academy of Sciences,
  Beijing 100190, China} \affiliation{Institute of
  Theoretical Physics, Chinese Academy of Sciences, Beijing 100080,
  China }
\author{Bing-Shen Wang} \affiliation{State
  Key Laboratory of Semiconductor Superlattice and Microstructure \\
  and Institute of Semiconductor, Chinese Academy of Sciences, Beijing
  100083, China\\}
 \author{Zhao-Bin Su} \affiliation{Institute of
  Theoretical Physics, Chinese Academy of Sciences, Beijing 100080,
  China } \affiliation{Center for Advanced Study, Tsinghua University,
  Beijing 100084, China}

\date{\today}
\begin{abstract} {The symmetry group analysis is applied to classify
    the phonon modes of $N$-stacked graphene layers (NSGL's) with AB- 
    and AA-stacking, particularly their infra-red and Raman properties. 
    The dispersions of various phonon modes are calculated in
    a multi-layer vibrational model, which is generalized from the
    lattice vibrational potentials of graphene to including the
    inter-layer interactions in NSGL's. The experimentally reported 
    red shift phenomena in the layer number dependence of the intra-layer 
    optical C-C stretching mode frequencies are interpreted. An interesting low frequency inter-layer optical mode is revealed to be Raman or Infra-red active in even or odd NSGL's respectively.
  Its frequency shift is sensitive to the layer
    number and saturated at about 10 layers.}
\end{abstract}

 \maketitle
\section{Introduction}
In recent years, the ultrathin graphite films, i.e. $N$ ($>1$) stacked
graphene layers (NSGL's), are successfully fabricated.\cite{Novoselov1,
  Berger} Extensive studies have been devoted to these systems
due to their competitive capability for the design of novel
nano-devices. It's desirable to carry out a through symmetry analysis to figure
out the corresponding electron and phonon spectra, and to reveal the
relevant selection rules and the optical activities. Moreover, to our
knowledge, although there are some studies on the electronic
structure,\cite{Peeters1} no existing works concern the phonon dispersions
for the NSGL's. Recently it is reported that the
frequencies of optical C-C stretching mode in the NSGL's decrease with
increasing $N$.\cite{Ferrari, Gupta, Anindya} The amplitude of this
red shift is about $3 - 5$, $5 - 6$, and 8~cm$^{-1}$ in these three
experiments respectively.  Theoretical explanation of this red shift
is required.

In this paper, the symmetry analysis is referred for both the AB- and
AA-stacked lattice structures, while the latter has the point group
$D_{6h}$ irrespective of the even-oddness of $N$.  We classify the
phonon normal modes at $\Gamma$ point and determine their Raman and
Infra-red (Ir) properties. We
generalize the force constant model of graphene\cite{Aizawa} into
NSGL's and it can be applied to calculate the phonon dispersions for the NSGL's in AB- or
AA-stacking with arbitrary layer number $N$. 

For the intra-layer
optical C-C stretching mode with frequency around 1600~cm$^{-1}$, it 
is Raman active for all NSGL's and the calculated frequencies exhibit
layer-number dependence as that the frequency decreases with $N$
increasing. The red shift values for the AB- (AA-) stacked systems are
about 2~cm$^{-1}$ (4~cm$^{-1}$) which are in consistent with the
experimental measurements. In the medium frequency range around 800~cm$^{-1}$,
the out-of-plane optical mode is Ir active for AB-stacked structure but
neither Raman nor Ir active in AA-stacking. Its frequency behaviors a blue
shift as layer number increasing. There is an interesting inter-layer optical
mode in the low frequency region which is Raman active in the NSGL's with $N$ even (ENSGL's) while 
Ir active in NSGL's with $N$ odd (OENSGL's). Its
frequency value depends on the layer number $N$ more sensitively and
increases from 106~cm$^{-1}$ (94.5~cm$^{-1}$) to 149.8~cm$^{-1}$
(133.6~cm$^{-1}$) for the AB- (AA-) stacked NSGL's, which is of an
order less than those intra-layer optical modes. Phonon dispersions
for the AA-stacked 3-dimensional (3D) graphite are also discussed.

The present paper is organized as follows. In Sec.~II, the lattice
configuration is illustrated for the NSGL's.  Sec.~III is devoted to
the symmetry analysis for the phonon modes. The vibrational potential
energy is discussed in Subsec.~IV~A, while the main results and
relevant discussions on phonon spectrum calculations are presented in
Subsec.~IV~B. The paper ends with a brief summary in Sec.~V.

\section{lattice configuration}
\subsection{AB-stacked}
\begin{figure}
  \begin{center}
    \scalebox{1.2}[1.4]{\includegraphics[width=7cm]{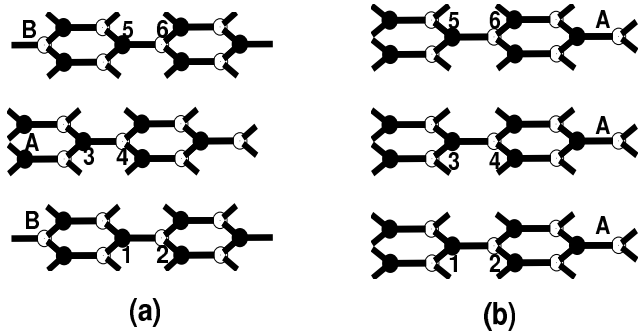}}
  \end{center}
  \caption{The sketch of the configurations of AB-stacked in (a) and 
      AA-stacked in (b) for multi-layer graphene.}
  \label{Fig:unitcell}
\end{figure}

It is known that graphene is a single layer of carbon atoms with the
honeycomb lattice configuration which is characterized by the $D_{6h}$
symmetry.\cite{Dresselhaus} The 3D graphite is AB-stacked honeycomb
lattice, where the B layers are achieved by shifting the A layers
along one of its first-nearest carbon-carbon bonds in the horizontal
plane as shown in Fig.~\ref{Fig:unitcell}(a). The space group of the
3D graphite is non-symmorphic group $D_{6h}^{4}$ with non-primitive
translation $ \vec\tau=\frac{1}{2}\vec{c}$ (primitive translation
$\vec t=n_1 \vec a_1+n_2 \vec a_2+ n_3\vec c$).\cite{Brillson} The
distance between two adjacent layers is about $\frac{c}{2}=3.35\AA$
which is much larger than the bond length between two nearest-neighbor
atoms in the plane, $b=1.42\AA$.

The NGSL's are also constructed by AB-stacked honeycomb lattice, but
with limited number of layers. Although the structures of each layer of
3D graphite and NSGL's are the same, the corresponding symmetry groups
are different since the displacement symmetry along $\vec{c}$ axis no
longer exists for NSGL's, so does the symmetry associated with
$\vec\tau$. Now the symmetry group becomes a direct product of a 3D
point group and a 2D translational group. And the point groups are
different for ENSGL's and ONSGL's as mentioned in
Ref.~\onlinecite{Manes}. For the ENSGL's, a center of inversion
$\sigma_{i}$ occurs in the middle of atom 4 in the $\frac {N}{2}$-th
layer and 5 in the $\frac {N}{2}+1$-th layer as shown in
Fig.~\ref{Fig:unitcell}(a). There are one 3-fold main axis in the
direction perpendicular to the layers and three 2-fold axes,
$C_{2}''$, perpendicular to the main axis and at angles of $\pi/3$ to
each other. All these symmetry operations together constitute the
point group $D_{3d}=\{E, 3C_{3}, 3C_{2}''\}\times\{E, \sigma_{i}\}$.
In the ONSGL's, instead of a center of inversion there is a reflection
symmetry $\sigma_{h}$ with the middle layer as its reference plane. A
3-fold main axis exists in the direction of $z$-axis and three 2-fold
axes, $C_{2}'$, are perpendicular to it.  We notice that these three
2-fold axes $C_{2}'$ are one to one perpendicular to those of
$C_{2}''$. Consequently, the symmetry group for the ONSGL's is
$D_{3h}=\{E, 3C_{3}, 3C_{2}'\}\times\{E, \sigma_{h}\}$.

The environments for an atom in the graphite and NSGL's are different
from that in 2D graphene. For each carbon atom in the graphene layer,
there are three nearest-neighbor carbon atoms and six next-nearest-
neighbors. There are four carbon atoms 1, 2, 3, 4 in a unit cell of
graphite, as represented in Fig.~\ref{Fig:unitcell}(a). For atom 4 in
A layer, there are two inter-layer nearest-neighbors in each of the
two adjacent layers, with the distance $c$. Also, in each of the two
adjacent layers, there are three inter-layer next-nearest-neighbor
atoms around atom 4 with distance $\sqrt{b^{2}+(\frac{c}{2})^{2}}$. As illustrated
in Fig.~\ref{Fig:unitcell}(a), the adjacent environment for atom 3 is
quite different from that of atom 4. It has no inter-layer
first-neighbors with the same distance as $\frac{c}{2}$. However, atom 3 has six
inter-layer neighbors in each of the two adjacent layers with the same
distance as that of the second-nearest-neighbor of atom 4.
\subsection{AA-stacked}
The AA-stacked NSGL's (AA-NSGL's) are constructed by AA-stacked
honeycomb lattice where all layers have the same configuration. In the
AA-stacked system, the ENSGL's, ONSGL's and the 3D graphite have the
same point group $D_{6h}$ which is the symmetry of the graphene.  As
shown in Fig.~\ref{Fig:unitcell}(b), the environments for carbon
atoms in the AA-stacked NSGL's are quite different from those in AB-stacked
systems. For each atom, there are two inter-layer nearest-neighbors in
each of the two adjacent layers with the distence $\frac{c}{2}$. And each atom
has six inter-layer second-nearest-neighbors with distance
$\sqrt{b^{2}+(\frac{c}{2})^{2}}$ in its two adjacent layers. We notice here that
in the AA-stacked 3D graphite, there are only two atoms in the unit
cell and the primitive translation along $c$-axis is $\vec{c}/2$, which
is half of the correspondence in the AB-stacked 3D graphite.

\section{symmetry analysis for the phonon modes}
The dynamical representation
$\Gamma^{dyn}=\Gamma^{v}\bigotimes\Gamma^{atom}$ can be decomposed
into the irreducible representations of the symmetry group, with the
lattice displacements as the bases, where $\Gamma^{v}$ is the vector
representation and $\Gamma^{atom}$ is the permutation representation
of the group. By applying the projection operator technique, we carry
out the decomposition of the dynamical representation into the
irreducible representations for the NSGL's with $N$ even and odd
respectively. According to Elliott,\cite{Elliott} the Ir
active phonon modes belong to the irreducible representations
decomposed from the vector representation $\Gamma^{v}$, while the
Raman active modes correspond to the irreducible representations shown
up in the decomposition of a six-dimensional representation with bases
as the quadratic forms: $x^{2}+y^{2}$, $z^{2}$, $x^{2}-y^{2}$, $xy$,
$yz$, and $zx$. The three acoustic modes with zero frequency at the
$\Gamma$ point, which correspond to the vector representation
$\Gamma^{v}$, are excluded in the consideration of Ir and Raman active
modes. For comparison the corresponding results for the graphene and
3D graphite are also listed in the following. The symbol for the
irreducible representations we used here is the notation used in
Ref.~\onlinecite{Eyring} which is the most commonly used in the
treatment of molecules.
\begin{table*}[t]
  \caption{The symmetry analysis for the phonon modes at the
    $\Gamma$ point of the NSGL's with AA- or AB-stacking.
    Phonon modes are classified by the irreducible
    representations of $\Gamma^{dyn}$ in the fourth column.
    The irreducible representations of the Ir and Raman
    active modes are listed in the fifth and sixth column respectively.}
  \label{Tab:SymmetryAnalysis}
  \begin{ruledtabular}
    \begin{tabular}{|l|c|c|c|c|c|}
      &&group&$\Gamma^{dyn}$&$\Gamma^{Ir}$&$\Gamma^{R}$\\
      \hline
      graphene\cite{Dresselhaus}&&$D_{6h}$&$A_{2u}\bigoplus B_{2g}\bigoplus E_{1u}\bigoplus E_{2g}$&/&$E_{2g}$\\
      \hline &ENSGL's&$D_{3d}$\cite{Manes}&$N(A_{1g}\bigoplus A_{2u}\bigoplus E_{g}\bigoplus E_{u})$&$(N-1)A_{2u}\bigoplus (N-1)E_{u}$&$NA_{1g}\bigoplus NE_{g}$\\
      AB-stacked&ONSGL's&$D_{3h}$\cite{Manes}&$(N-1)A_{1}'\bigoplus (N+1)A_{2}''\bigoplus (N+1)E'$&$NA_{2}''\bigoplus N E'$&$(N-1)A_{1}'\bigoplus NE'\bigoplus (N-1)E''$\\
      &&& $\bigoplus (N-1)E''$ &&\\
      &3D\cite{Mani}&$D_{6h}^{4}$&$2(A_{2u}\bigoplus B_{2g}\bigoplus
      E_{1u}\bigoplus
      E_{2g})$&$A_{2u}\bigoplus E_{1u}$&$2E_{2g}$\\
      \hline
      &ENSGL's&$D_{6h}$&$\frac{N}{2}(A_{1g}\bigoplus A_{2u}\bigoplus B_{1u}\bigoplus B_{2g}\bigoplus E_{1u}$&$(\frac{N}{2}-1)(A_{2u}\bigoplus E_{1u})$&$\frac{N}{2}(A_{1g}\bigoplus E_{1g}\bigoplus E_{2g})$\\
      &&& $\bigoplus E_{1g}\bigoplus E_{2g}\bigoplus E_{2u})$ &&\\
      AA-stacked&ONSGL's&$D_{6h}$&$\frac{N-1}{2}(A_{1g}\bigoplus B_{1u}\bigoplus E_{1g}\bigoplus E_{2u})$&$\frac{N-1}{2}(A_{2u}\bigoplus E_{1u})$&$\frac{N-1}{2}(A_{1g}\bigoplus E_{1g})\bigoplus \frac{N+1}{2}E_{2g}$\\
      &&& $\bigoplus \frac{N+1}{2}(A_{2u}\bigoplus B_{2g}\bigoplus E_{1u}\bigoplus E_{2g})$ &&\\
      &3D&$D_{6h}$&$A_{2u}\bigoplus B_{2g}\bigoplus E_{1u}\bigoplus E_{2g}$&/&$E_{2g}$\\
    \end{tabular}
  \end{ruledtabular}
\end{table*}

The symmetry analysis for the phonon modes and the Raman and Ir modes
are classified in Table.~\ref{Tab:SymmetryAnalysis}. In the
AB-stacked NSGL's, since the $\sigma_{i}$ symmetry and $\sigma_{h}$
symmetry can not coexist in the ENSGL's or ONSGL's, we can see two
straightforward consequences from the above classification for the Ir
and Raman active modes. Firstly, in the ENSGL's, phonon modes can not
be Ir and Raman active simultaneously (which is also true for the
gaphene and graphite). However, in the ONSGL's, the $N$ $E'$ modes are
both Ir and Raman active. This is because there is no inversion center
in the ONSGL's. Secondly, among the optical modes with their
vibrational displacements perpendicular to the constituent layers,
there is an exotic mode oscillating with each layer as a whole but
alternatively from layer to layer. It belongs to the $A_{1g}$ in the
ENSGL's and $A''_{2}$ in the ONSGL's. Since the $\sigma_{h}$ operation
exists only in the ONSGL's, this mode ($\omega_{1}$ mode) is Raman
active in the ENSGL's while Ir active in the ONSGL's.

In the AA-stacked NSGL's with $N$ even or odd, the symmetry group is
$D_{6h}$ which includes both $\sigma_{i}$ and $\sigma_{h}$. As a
result, phonon modes can not be Ir and Raman active simultaneously.
The $\omega_{1}$ mode mentioned above belongs to the $A_{1g}$ in the
ENSGL's and $A_{2u}$ in the ONSGL's. This mode is Raman active in the
ENSGL's while Ir active in the ONSGL's which is the same as the
AB-stacked NSGL's. Nevertheless, its vibrational mode favors to take
the maximum advantage of the inter-layer interactions. It would be
sensitive and useful experimentally to identify the even-oddness of
the NSGL's with a few layers.

\section{calculation for the phonon dispersion}
\subsection{Vibrational potential energy}
The vibrational potential energy for a graphene sheet can be described
by five quadratic terms with the rigid rotational symmetry
implemented.\cite{Aizawa, Jiang2} They are the 1st and 2nd
nearest-neighbor stretching, the in-plane bond angle variations, the
out-of-surface bond bending and the bond twisting energies. From the
modality of atomic movements, we can also classify the inter-layer
vibrational potential terms into three types: The first one describes
the stretching movements between the two atoms located in the adjacent
layers. The second describes the relative movement between the two
pairs of atoms with a common one as an apex. That is this type of
movements involves three atoms to form one bond in a layer and another
connecting the two nearest layers. The third involves more than three
atoms according to the specific bond configurations. As shown in
Fig.~\ref{Fig:unitcell}, there is only one inter-layer
nearest-neighbor carbon-carbon bond in each unit cell (the bond
between atoms 1 and 4). So that just the twisting potential on the
inter-layer bond is encountered here. The whole of these terms is
actually a modified valence force field model to account some interactions for far away atoms
in response to the bond charge effect in certain extent. Since the
inter-layer bonds are much longer than that in the plane, all above three 
type interactions are one or two orders
less than their counterparts in layer and they
themselves have comparable contributions. In the following, the
inter-layer terms are written in the AB-stacked system and they can be
similarly generalized to the AA-stacked system.

\begin{table}[t]
  \caption{Comparison of several mode frequencies (in the unit of
    cm$^{-1}$) for the AB-stacked 3D graphite between the our calculation results and the
    experimental values.\cite{Nicklow, Maultzsch}}
  \label{Tab:Fit}
  \begin{ruledtabular}
    \begin{tabular}{cllll}
      Reps &$A_{1}^{'}$ & $E_{2g}$ & $A_{2u}$ & $E_{2g}$ \\
      \hline
      experiments& 30\cite{Nicklow} & 40\cite{Nicklow} & 868\cite{Maultzsch} & 1586\cite{Maultzsch}\\
      theory &30.2 & 42.7 & 869.9 & 1586.6 \\
    \end{tabular}
  \end{ruledtabular}
\end{table}

(1). The inter-layer bond stretching energies $V^{(int)}_l
(V^{(int)}_{sl})$ have the form as:
\begin{eqnarray}
  \sum_{i,j}\frac{\hat{k}_l}{2}[(\vec{u}_{i}-\vec{u}_{j})\cdot
  \vec{e}_{ij}^{l}]^{2},
  \label{Eq:Potential1}
\end{eqnarray}
where $\vec{u}_{i}$ $(\vec{u}_{j})$ is the displacement vector of the
atom $i$ $(j)$ and $\vec{e}_{ij}^{l}$ is the unit vector from atom $i$
to atom $j$. If the summation is taken over the nearest-neighbored
inter-layer pair of atoms, the corresponding force constant is denoted
as $\hat{k}_l$ while for next nearest-neighbor inter-layer pairs we
have the force constant as $\hat{k}_{sl}$.

(2). For the three atoms 1, 4 and $i$, where $i$ is the in-plane
nearest neighbor of atom 1 (see Fig.~\ref{Fig:unitcell}), we found
that under a specific configuration with atom $i$ rather than atom 1
as an apex, while the force being along the corresponding bond
direction instead of perpendicular direction, a correlation term
$\hat{k}_{rr}$ has the most and sensitive contribution to the layer
dependence of the inta-layer C-C stretching optical modes,
\begin{eqnarray}
  \frac{\hat{k}_{rr}}{2}\sum_{i}[(\vec{u}_{1}-\vec{u}_{i})\cdot
  \vec{e}_{i1}^{l}-(\vec{u}_{4}-\vec{u}_{i})\cdot
  \vec{e}_{i4}^{l}]^2. \nonumber
\end{eqnarray}
Actually the two square terms in above modality have already been
accounted in the in-plane and inter-plane stretching terms
respectively. Only the across term is left,
\begin{eqnarray}
  V_{rr}=-\hat{k}_{rr}\sum_{i}[(\vec{u}_{1}-\vec{u}_{i})
  \cdot \vec{e}_{i1}^{l}]
  [(\vec{u}_{4}-\vec{u}_{i})\cdot \vec{e}_{i4}^{l}]\; ,
  \label{Eq:Potential4}
\end{eqnarray}
which weakens the interaction between two adjacent layers.  The
positive definite condition for getting real frequencies is
$\hat{k}_{sl}\geq \hat{k}_{rr}$.

(3). The twisting potential for an inter-layer bond between atoms 1
and 4 is coming from the two sets of three nearest-neighbors of atoms
1 and 4 respectively. It can be described as
\begin{eqnarray}
  V_{tw}=\frac{\hat{k}_{tw}}{2}[\sum_{i}(\vec{u}_{i}-\vec{u}_{1})\cdot \vec{e}_{i}^{\theta}
  -\sum_{j}(\vec{u}_{j}-\vec{u}_{4})\cdot
  \vec{e}_{j}^{\theta}]^{2},
  \label{Eq:Potential3}
\end{eqnarray}
where $\sum_{i}$ and $\sum_{j}$ represent the summation over the three
intra-plane first-nearest-neighbors for atoms 1 and 4
respectively. $\vec{e}_{i}^{\theta}=\vec{e}_{z}\times
\vec{e}_{1i}^{l}$ is the tangential unit vector in the plane formed by
three atoms 1, 4, and $i$. The expression in quadratic form as a whole
ensures a proper definition for the torsion angle. For pure rotations
around the bond, this expression gives zero torsion consistently. In
contrast, the bond is most severely twisted when the three neighbors
around atom 1 and those of atom 4 rotate reversely.

We stress here that, all of the above four inter-layer vibrational
potential energy terms satisfy the rigid rotational symmetry
requirements\cite{Popov, Mahan, Jiang2} which guarantees the existence
of the flexure modes in the low dimensional systems. Although we establish the vibrational
potential terms based on the analysis to the modality of movements,
the bond charge effect especially along the perpendicular direction
has been involved by extending the valence force field beyond the nearest
neighbors. Comparing, for example, the above $\hat{k}_{rr}$ term with
that $V_{b-b}$ in Ref.~\onlinecite{Mahan2}, which
is followed from the bond-charge model, they have the same negative
cross term.

\subsection{Results and discussion}

\begin{figure}
  \begin{center}
    \scalebox{1.0}[1.0]{\includegraphics[width=7.6cm]{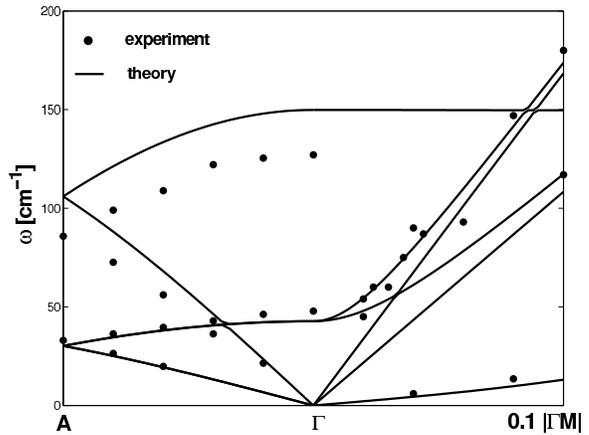}}
  \end{center}
  \caption{Phonon dispersion for the 3D graphite in the low-frequency
    region.  Solid dots are the experimental results of
    Ref.~\onlinecite{Nicklow}.  Our theoretical calculations are shown
    in lines.}
  \label{Fig:3DFitLow}
\end{figure}
\begin{figure}
  \begin{center}
    \scalebox{1.0}[1.0]{\includegraphics[width=7.6cm]{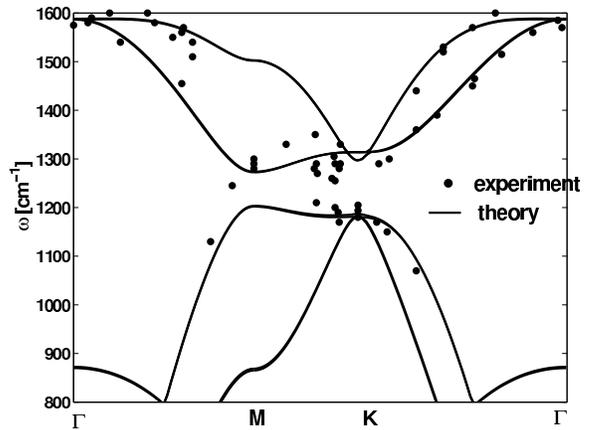}}
  \end{center}
  \caption{Phonon dispersion for the 3D graphite in the high-frequency
    region.  Solid dots are the experimental results.\cite{Maultzsch,
      Mohr} In Refs.~\onlinecite{Maultzsch, Mohr}, those phonon wave
    vectors $\vec{q}$, which were not exactly along the ~$\Gamma$-M or
    ~$\Gamma$-K-M direction, were projected onto the closest
    high-symmetry direction.  Lines are our theoretical calculations.}
  \label{Fig:3DFitHigh}
\end{figure}

The five intra-layer force constants we used in the following are
taken from Ref.~\onlinecite{Jiang} with a minor modification. We
adjust the four inter-layer force constants to fit the experimental
values of four modes in 3D graphite as shown in Table~\ref{Tab:Fit}.
The fitting error for phonon modes is kept less than $7\%$. The
inter-layer force constants are then fitted as
$\hat{k}_{l}=0.77$~Nm$^{-1}$, $\hat{k}_{sl}=0.95$~Nm$^{-1}$,
$\hat{k}_{tw}=0.64$~Nm$^{-1}$, $\hat{k}_{rr}=0.9$~Nm$^{-1}$.

Base on the above fitted vibrational potential energy with nine terms,
we calculate the dispersion curves for the AB-stacked graphite. As
illustrated in Figs~\ref{Fig:3DFitLow} and ~\ref{Fig:3DFitHigh}, our
theoretical calculations meet the experimental results not only in the
low frequency \cite{Nicklow} but also in the high frequency regions.
\cite{Maultzsch, Mohr} The excellent consistency with the experimental data
shows that our model and parameters are reasonable and applicable.

\begin{table}[t]
  \caption{The Raman and Ir mode frequencies (in the unit of cm$^{-1}$)
    for the AB-stacked 3D graphite, AB-stacked 2-layer and AA-stacked 3D grphite are listed. 
    The irreducible representations
    are presented in the brackets following the frequency values.}
  \label{Tab:Raman}
  \begin{ruledtabular}
    \begin{tabular}{|l|cc|cc|}
      &Raman&&Infra-red&\\
      \hline
      AB- 3D & 42.7 ($E_{2g}$) & 1586.7 ($E_{2g}$) & 869.9 ($A_{2u}$) & 1588.2 ($E_{1u}$)\\
      \hline
      AB- & 30.2 ($E_{g}$) & 106 ($A_{1g}$) & 868.7 ($A_{2u}$) & 1588.1 ($E_{u}$)\\
      2-layer & 867.4 ($A_{1g}$) & 1587.3 ($E_{g}$)\\
      \hline
      AA- 3D & 1584.7 ($E_{2g}$) &  & & \\
    \end{tabular}
  \end{ruledtabular}
\end{table}

\begin{figure}
  \begin{center}
    \scalebox{1.0}[1.0]{\includegraphics[width=7.6cm]{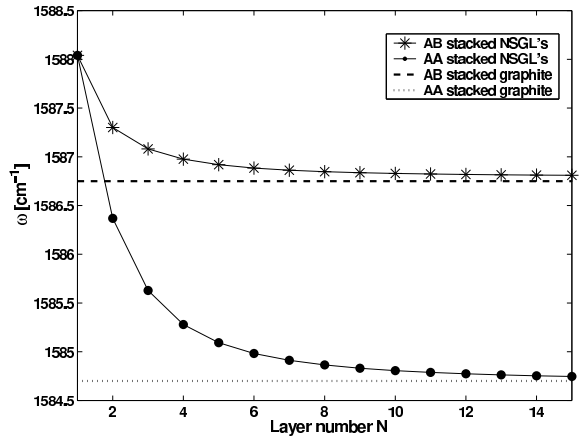}}
  \end{center}
  \caption{The frequency value for the optical C-C stretching mode vs
    the layer number $N$.  Lines are draw to guide eyes.}
  \label{Fig:ModeIntraLayer}
\end{figure}
\begin{figure}
  \begin{center}
    \scalebox{1.0}[1.0]{\includegraphics[width=7.6cm]{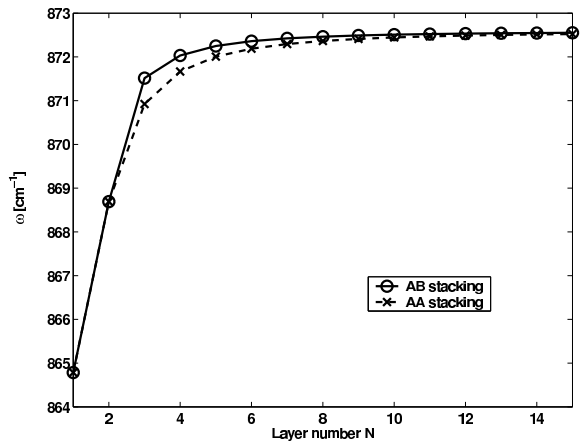}}
  \end{center}
  \caption{The frequency value for the out-of-plane optical mode vs
    the layer number $N$. This mode is Ir active in the AB stacking while 
    it is neither Ir nor Raman active in the AA stacking. Lines are draw to guide eyes.}
  \label{Fig:ModeOut}
\end{figure}
\begin{figure}
  \begin{center}
    \scalebox{1.1}[1.2]{\includegraphics[width=7.6cm]{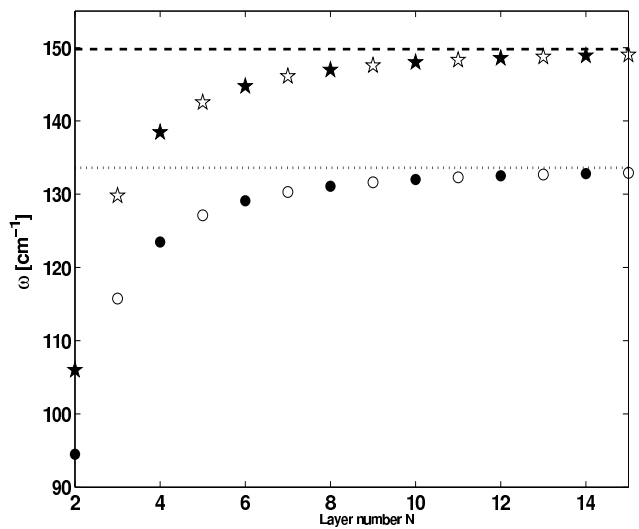}}
  \end{center}
  \caption{The frequencies of the inter-layer optical mode vs the layer
    number $N$. Datas for the AB and AA stacked NSGL's are designated by 
    pentagrams and circles, respectively. The Raman and Infra-red activities 
    for this mode are displayed by the full and empty symbols, respectively.
    The broken and dashed lines correspond to the frequencies for the AB-stacked 
    and AA-stacked graphite, respectively.}
  \label{Fig:ModeInterLayer}
\end{figure}
\begin{figure}
  \begin{center}
    \scalebox{1.0}[1.0]{\includegraphics[width=7.6cm]{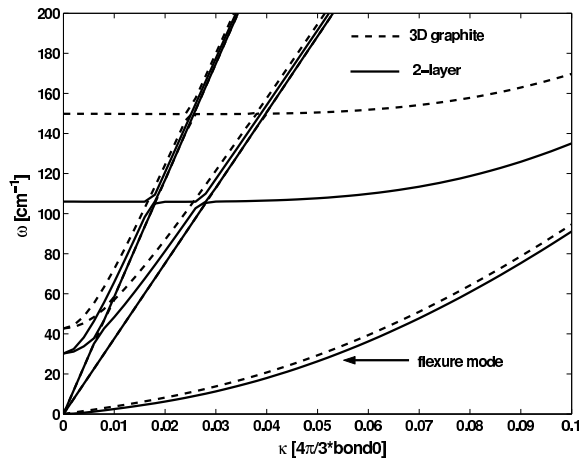}}
  \end{center}
  \caption{In the low frequency region, there is significant
    difference between 3D graphite and the 2-layer graphene.}
  \label{Fig:2Layer3DLow}
\end{figure}

With the above force constants, we can calculate the phonon dispersion
for NSGL's of the AA- or AB-stacking with an arbitrary layer
number $N$. In Fig.~\ref{Fig:ModeIntraLayer}, the calculated frequency
of the intra-layer optical C-C stretching mode is represented with
different stacked styles and layer number $N$. The layer dependence of
the frequency shows up a red shift behavior which is in agreement with
the experimental measurements. The frequency value for this mode is
about 1588~cm$^{-1}$ in the single graphene layer and decreases with
increasing $N$ and almost saturates at $N=10$. The limit is
1586.7~cm$^{-1}$ (1584.7~cm$^{-1}$) in the AB- (AA-) stacked system
respectively. The amount of red shift value in our calculation
corresponds excellently with that measured by experiments within the ranges $3 - 5$, $5 -
6$, and 8~cm$^{-1}$ in Refs.~\onlinecite{Ferrari}, \onlinecite{Gupta}
and \onlinecite{Anindya}, respectively.

The out-of-plane optical mode, belonging to the $A_{2u}$ ($B_{2g}$)
irreducible representation in the AB (AA) stacking, is Ir active in the AB-stacking
yet inactive in the AA-stacking irrespective of the even-oddness of the 
layer number $N$ and is useful in determining whether the NSGL's is of AB or AA stacking. 
As shown in Fig.~\ref{Fig:ModeOut}, frequencies for 
this mode depend on the layer number $N$ and increase from 864.8~cm$^{-1}$
to 872.6~cm$^{-1}$ in both the AB and AA stacking. In contrast to the C-C 
stretching optical mode, this mode frequency exhibits a blue shift type
layer dependence which could be identified with the development of the experimental
technique.

For the inter-layer optical mode, the layer number dependence of the
frequency value is shown in Fig.~\ref{Fig:ModeInterLayer}. This mode
takes the greatest advantage of the inter-layer interation and is
considerably dependent on the layer number $N$ and the stack style AB
or AA. In case of $N=2$, the $\omega_{1}$ mode has the frequency values
106~cm$^{-1}$ and 94.5~cm$^{-1}$ for the AB- and AA-stacked NSGL's
respectively. The frequencies of the $\omega_1$ mode increase with
increasing $N$ and almost come to the limit values at $N=10$. The limit
values are 149.8~cm$^{-1}$ and 133.6~cm$^{-1}$ for the AB- and
AA-stacked NSGL's respectively. The frequency differences as well as the
Raman versus Ir (see Sec. III) of the $\omega_{1}$ mode in NSGL's with
different layers might inspire considerably experimental interest in
the $\omega_1$ mode.

We then calculate the phonon dispersion for the 2-layered AB stacking, in
comparison with that of the 3D graphite. The most significant
difference between the 2-layer graphene and the 3D graphite lies in
the low-frequency region around the $\Gamma$ point as shown in
Fig.~\ref{Fig:2Layer3DLow}. The frequencies of the low-frequency
optical modes in the 2-layer graphene are much smaller than their
counterparts in the 3D graphite. The frequencies of the Raman and Ir
active modes are shown in the third line of Table~\ref{Tab:Raman}
among which the two $A_{1g}$ modes have the frequency value
$\omega_{1}$=106~cm$^{-1}$ and $\omega_{2}$=867.4~cm$^{-1}$. In fact,
$\omega_{1}$ and $\omega_{2}$ modes are the above mentioned $NA_{1g}$
modes of the ENSGL's specified to $N=2$.

\begin{figure}
  \begin{center}
    \scalebox{1.0}[1.0]{\includegraphics[width=7.6cm]{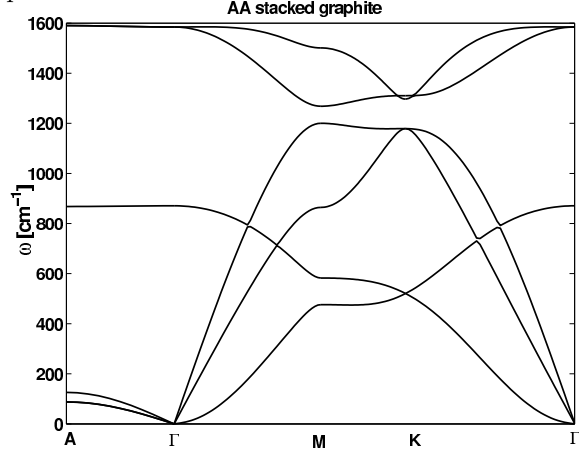}}
  \end{center}
  \caption{The phonon dispersion along some high-symmetry directions
    for the AA-stacked 3D graphite. There are only six branches in the
    figure, since the unit cell in the AA-stacked 3D graphite contains
    two atoms.}
  \label{Fig:AA3Dgraphite}
\end{figure}

We further calculate the phonon dispersion curve for the AA-stacked 3D
graphite as shown in Fig.~\ref{Fig:AA3Dgraphite}. Since the unit cell
contains only two atoms in contrast to that of the AB-stacked
graphite, there are six branches of phonon dispersion. Along $\Gamma A$ in
the Brillouin zone, the lowest and highest branches which correspond
to the in-plane acoustic and optical vibrational modes are doubly
degenerate, while the remaining two branches describing the
out-of-plane vibration are non-degenerate. At the $A$ point in
the Brillouin zone, there is a phase factor difference of $\pi$
between two adjacent layers for the out-of-plane motion, among which
the $\omega_{1}$ mode has the frequency value of 133.6~cm$^{-1}$. In
the fourth line of Table~\ref{Tab:Raman}, the Raman and Ir active
modes for the AA-stacked 3D graphite are listed which are to be
confirmed in future experiments.

\section{conclusion}
Based upon a thorough investigation of the lattice symmetry of the
NSGL's, the Raman and Ir properties, in particular its layer dependence, is
systematically studied. With a proposed generalized vibrational
potential, we further calculate the phonon dispersion of various modes
of the AB- or AA-stacked NSGL's where the layer dependence is also
stressed. The calculated frequencies of optical C-C stretching mode 
exhibit a red shift as layer number increasing in both the AB- and 
AA-stacked NSGL's, and the shift value 2~cm$^{-1}$ (4~cm$^{-1}$) for 
AB- (AA-) stacked NSGL's is in good consistent with the experimental measurements. 
The out-of-plane optical mode with frequency around about 800~cm$^{-1}$ is 
Ir active in AB-stacked structure yet neither Raman nor Ir active
in AA-stacking. Its frequency shows a blue shift layer dependence.
We also predict that the frequency of the inter-layer optical mode increases with $N$
increasing. Since this mode is more sensitive to the layer number $N$, 
it should be experimentally interesting in
determining the lattice structure properties of the NSGL's.


\begin{thebibliography}{}
\bibitem{Novoselov1} K. S. Novoselov, A. K. Geim, and S. V. Morozov
  \textit{et al.}, Science \textbf{306}, 666 (2004).

\bibitem{Berger} C. Berger, Z. Song, and T. Li \textit{ et al.},
  J. Phys. Chem. B \textbf{108}, 19912 (2004).

\bibitem{Peeters1} B. Partoens and F. M. Peeters, Phys. Rev. B
  \textbf{74}, 075404 (2006); Phys. Rev. B \textbf{75}, 193402 (2007).

\bibitem{Ferrari} A. C. Ferrari, J. C. Meyer, and V. Scardaci
  \textit{et al}., Phys. Rev. Lett. \textbf{97}, 187401 (2006).

\bibitem{Gupta} A. Gupta, Gugang Chen, and P. Joshi \textit{et al}.,
  Nano Lett., \textbf{6}, 2667, (2006).

\bibitem{Anindya} Anindya Das, Biswanath Chakraborty, and A. K. Sood,
  arXiv:cond-mat/0710.4610v1.

\bibitem{Aizawa} T. Aizawa, R. Souda, S. Otani, Y. Ishizawa, and
  C. Oshima, Phys. Rev. B \textbf{42}, 11469 (1990); \textbf{43},
  12060(E) (1991).

\bibitem{Dresselhaus} R. Saito, G. Dresselhaus, and M.S. Dresselhaus,
  \textit{Physical Properties of Carbon Nanotubes} (Imperial College
  Press, 1998).

\bibitem{Brillson} L. Brillson, E. Burskin, A. A. Maradudin, and
  T. Stark, \textit{The Physics of Semimetals and Narrow-Gap
    Semiconductors}, Ed. D. L. Carter and R. T. Bate, (Pergamon Press,
  London/Oxford 1971).

\bibitem{Manes} J. L. Manes, F. Guinea, and Maria A. H. Vozmediano,
  Phys. Rev. B \textbf{75}, 155424 (2007).

\bibitem{Elliott} J. P. Elliott and P. G. Dawber, \textit{Symmetry in
    Physics, Vol. 1}, (Macmillan Press, Ltd., 1979).

\bibitem{Eyring} Eyring, Walter, and Kimball, \textit{Quantum
    Chemistry}, John Wiley and Sons, (Inc., New York, 1940).

\bibitem{Mani} K. K. Mani and R. Ramani, Phys. Status Solidi B
  \textbf{61}, 659 (1974).

\bibitem{Jiang2} J. W. Jiang, H. Tang, B. S. Wang, and Z. B. Su,
  Phys. Rev. B \textbf{73}, 235434 (2006).

\bibitem{Popov} V.N. Popov, V.E. Van Doren, and M. Balkanski,
  Phys. Rev. B \textbf{61},3078( 2000).

\bibitem{Mahan} G.D. Mahan, and Gun Sang Jeon, Phys. Rev. B
  \textbf{70}, 075405 (2004).

\bibitem{Mahan2} Gun Sang Jeon, and G.D. Mahan, Phys. Rev. B
  \textbf{72}, 155415 (2005).

\bibitem{Jiang} Jin-Wu Jiang, Hui Tang, Bing-Shen Wang and Zhao-Bin
  Su, J. Phys.: Condens. Matter \textbf{20}, 045228 (2008) .


\bibitem{Nicklow} R. Nicklow, N. Wakabayashi, and H. G. Smith,
  Phys. Rev. B \textbf{5}, 4951 (1972).

\bibitem{Maultzsch} J. Maultzsch, S. Reich, and C. Thomsen \textit{et
    al}., Phys. Rev. Lett. \textbf{92}, 075501 (2004).

\bibitem{Mohr} M. Mohr, J. Maultzsch, and S. Reich \textit{et al}.,
  Phys. Rev. B \textbf{76}, 035439 (2007).
\end{thebibliography}
\end{document}